\documentclass[12pt,a4paper]{article}
\usepackage{graphicx}
\renewcommand{\baselinestretch}{1.5}
\newcommand{\beq}{\begin{equation}}
\newcommand{\eeq}{\end{equation}}
\newcommand{\ds}{\displaystyle}
\newcommand{\form}{\mbox{C$_{12}$EO$_{6} \,$}}
\newcommand{\formeau}{\mbox{C$_{12}$EO$_{6}$/H$_{2}$O$\,$}}
\newcommand{\dgr}{ {\,}^{o} \mbox{C}}

\begin{document}

\date{}
\title{Connectivity of the Hexagonal, Cubic, and Isotropic Phases of the \formeau Lyotropic Mixture Investigated by Tracer Diffusion and X-ray Scattering}

\author{D. Constantin\protect\footnote{Author for correspondence. E-mail address : dcconsta@ens-lyon.fr; Tel : +0033-4 72 72 83 75; Fax : +0033-4 72 72 80 80} \  and P. Oswald \\  {\em Laboratoire de Physique de l'ENS de Lyon,} \\ {\em 46 All\'ee d'Italie, 69364 Lyon Cedex 07, France} \\   \\  M. Imp\'eror-Clerc, P. Davidson, and P. Sotta \\ {\em Laboratoire de Physique des Solides, Univ. Paris-Sud,} \\ {\em B\^atiment 510, 91405 Orsay Cedex, France} \\ }

\renewcommand{\baselinestretch}{1}

\maketitle

\renewcommand{\baselinestretch}{1.5}

\begin{abstract}

The connectivity of the hydrophobic medium in the nonionic binary
system \formeau is studied by monitoring the diffusion constants
of tracer molecules at the transition between the hexagonal
mesophase and the fluid isotropic phase. The increase in the
transverse diffusion coefficient on approaching the isotropic
phase reveals the proliferation of bridge-like defects connecting
the surfactant cylinders. This suggests that the isotropic phase
has a highly connected structure. Indeed, we find similar
diffusion coefficients in the isotropic and cubic bicontinuous
phases. The temperature dependence of the lattice parameter in the
hexagonal phase confirms the change in connectivity close to the
hexagonal--isotropic  transition. Finally, an X-ray investigation
of the isotropic phase shows that its structure is locally similar
to that of the hexagonal phase.

\end{abstract}


\section{Introduction}

The phase di\-a\-gram of the \formeau system was determined more
than 15 years ago \cite{mitchell}. Since then, extensive studies
have focused on the structure of the three mesophases it exhibits,
which are (on increasing surfactant concentration) hexagonal
$H_{\alpha}$, bicontinuous cubic $V_1$, and lamellar $L_ {\alpha}$
phases. Their characteristics are by now very well established
\cite{toledano}. The situation is less clear concerning the
organisation of the isotropic phase that borders all the
above-mentioned mesophases at higher temperature. The lack of
long-range order and of optical birefringence prevents the use of
common techniques such as X-ray diffraction or optical microscopy.
Therefore, the isotropic phase has been mostly studied by light
scattering and NMR, but the interpretation of experimental data is
largely model-dependent
\cite{nilsson,brown1,brown2,brown3,jonstromer}.

The isotropic phase in \formeau and other related systems has also
been studied by elastic and quasi-elastic neutron scattering
\cite{zulauf}. It is suggested from the observed fast relaxation
times that the isotropic phase is formed of small, globular
micelles, even in the vicinity of the hexagonal phase. However, no
direct evidence is presented to support this assertion.

Even at low concentration ($c < 30 \%$), the structure of the
isotropic phase is not yet completely clear. Interpretations in
terms of small, attractively interacting (clustering) micelles
have been put forward \cite{zulauf, triolo, magid, richtering}.
Other authors, however, claim that micelles grow in size and
become anisotropic with increasing temperature or concentration
\cite{brown1, cebula, ravey}.

It is therefore necessary to use complementary characterisation
methods. For instance, rheological investigations can provide
information about the structure of the isotropic phase
\cite{darrigo}.

Indirect geometrical information can be obtained by studying the
variation of structural properties of the surfactant aggregates
upon crossing a boundary from a liquid crys\-tal\-line phase to
the isotropic phase. The structure of the former, being better
known, can serve as reference. Furthermore, if the particular
property remains unchanged at the transition, one can assume that
the local structure in the higher-temperature phase is similar in
some respect to that in the low-temperature ordered phase.

Such experiments have already been performed at the
hexagonal-isotropic transition, the loss of hexagonal order upon
approaching the transition being monitored by NMR and optical
birefringence methods \cite{sallen1,sallen2}. These experiments
have shown that a significant fraction of the surfactant molecules
form defects near the isotropic liquid, which can be column ends
(with spherical caps) or bridges connecting neighbouring columns
(fig.\ref{fig:defects}). It is clear that the nature of the
isotropic liquid will be different depending on the nature of the
defects. Indeed, column ends will give rise to isolated micelles
whereas bridges should rather announce an isotropic liquid formed
of cylinders highly connected to each other and randomly oriented.

In order to discriminate between the two possibilities, we
measured the diffusion coefficients of a hy\-dro\-pho\-bic probe
(tracer molecule) which is dissolved in the lyotropic mixture
(sections 2.1 and 3.1). Indeed, it is reasonable to assume (as
demonstrated by the experimental results below) that the tracer is
confined to the surfactant cylinders and spends very little time
in water. Hence, its diffusion must give valuable structural
information about the surfactant aggregates, as we have shown in a
recent investigation of the lamellar phase in the \formeau system
\cite{constantin}. Two transitions were investigated : the
$H_{\alpha} \! \rightarrow \! I$ at the azeo\-tropic point of the
hexagonal phase (50.0 \% \form weight concentration) and the $V_1
\! \rightarrow \! I$ (concentration 59.0 \%). The reason why we
chose the azeotropic point for the first transition is that the
mixture behaves like a pure compound  at this particular
concentration, {\em i.e.} the freezing range vanishes, which
allows us to approach the transition very closely \cite{sallen1}.

We also measured very precisely the lattice parameter of the
hexagonal phase by X-ray scattering and obtained additional
qualitative information from the X-ray scattering of the isotropic
phase as a function of temperature (sections 2.2 and 3.2). Results
are discussed in section 4 within a numerical model for the
diffusion of the probe molecules which takes into account the
lifetime of the defects. Conclusions are drawn in section 5.

\section{Experimental}

\subsection{Diffusion}

The surfactant was purchased from Nikko Ltd. and used without
further purification. We used ultrapure water from Fluka. The
mixture was carefully homogenized. The samples were prepared
between two parallel glass plates, with a spacing of 75 $\mu$m as
described in detail elsewhere \cite{sallen1}. The hexagonal phase
was then oriented by the directional solidification technique
presented in reference \cite{oswald1}. No orientation procedure
was needed for the cubic phase, the diffusion tensor being
isotropic.

We used a hy\-dro\-pho\-bic fluorescent dye
(NBD-di\-oc\-tyl\-amin) at a concentration of about $0.1$ w \%.
Its presence decreases by \mbox{$\sim 1 \dgr$} the transition
temperature $T_h$ at the azeotropic point of the hexagonal phase ,
but the freezing range remains negligible (about 0.1 $\dgr$).

Our method of measuring the diffusion constant is a variant of the
technique known as fluorescence recovery after photobleaching
(FRAP). The experimental setup was originally designed for the
study of thin liquid films \cite{bechhoefer}. We focused the
TEM$_{00}$ mode of a multimode Ar$^+$-ion laser (total power 70
mW) on the sample, bleaching a spot about 40 $\mu$m in diameter.
The intensity profile of the beam was approximately Gaussian.
Typical bleaching times were of the order of 5 seconds. About $0.5
\dgr$ below the transition temperature the power had to be
decreased to around 15 mW to avoid local melting of the sample.
The evolution of the fluorescence intensity profile (proportional
to the concentration of non-bleached molecules $c_{\rm n} (x,y,t)$
) was then monitored for one minute using a cooled CCD camera
(Hamamatsu C4742). It is however easier to write the equations for
the concentration of {\em bleached} molecules, $c(x,y,t)=c_{\rm
tot} - c_{\rm n}(x,y,t)$. We write the diffusion equation for the
hexagonal phase, with $D_{\|}$ and $D_{\bot}$ the diffusion
constants along and across the columns, respectively. In the
equation for the cubic or isotropic phase, $D_{\|}$ and $D_{\bot}$
both take the same value, $D_{C}$ or $D_{I}$ respectively. The
bleaching is considered uniform through the sample, so the
concentration obeys a two-dimensional diffusion equation (in the
plane of the sample); the $x$ axis is along the columns :
${\mathbf{e}}_x \, \| \, \mathbf{n}$.

\beq \label{diffeq} \frac{\partial c}{\partial t} =
D_{\|}\frac{\partial ^2 c}{\partial x^2} + D_{\bot}\frac{\partial
^2 c}{\partial y^2} +\beta I_0 c_{\rm n} \eeq

\noindent where the source term $\beta I_0 c_{\rm n}$ accounts for
the bleaching due to the observation light, of homogenous
intensity $I_0$; its effect is negligible so we will ignore it
from now on.

The initial concentration profile is :

\beq \label{iniconc} c(x,y,0) = \frac{C}{a_1 a_2}  \exp \left ( -
\frac{x^2}{{a_1}^2}-\frac{y^2}{{a_2}^2}\right ) \eeq

\noindent where $a_1$ and $a_2$ are the semi-axes of the initial
spot. The time dependence reads :

\beq \label{timeconc} c(x,y,t) = C \frac{\exp \left (-
\frac{x^2}{{a_1}^2 + 4D_{\|}t} - \frac{y^2}{{a_2}^2 +
4D_{\bot}t}\right )}{\sqrt{({a_1}^2 + 4D_{\|}t)({a_2}^2 +
4D_{\bot}t)}} \eeq

\noindent In the hexagonal phase, where the diffusion is
anisotropic, the concentration profile is elliptical. The
diffusion coefficients are deduced from the images by fitting a
Gaussian function whose adjustable parameters are the two
coordinates of the center, the amplitude $C$, the two semi-axes
and the angle between the major axis of the ellipse and a given
axis in the plane.

\subsection{X-Ray Scattering}

High resolution X-ray scattering experiments were performed at the
H10 experimental station at the LURE synchrotron radiation
facility in Orsay, France \cite{gailhanou}. A wavelength $\lambda
= 0.155$ nm was selected by a two crystals monochromator. Harmonic
rejection was obtained by reflection of the X-ray beam on two
Rh-covered mirrors. The beam size was $ 0.5  \times 2 \, \mbox{mm}
^2$. The sample was set at the center of a Huber diffractometer
equipped with a crystal analyzer (Ge 111) . The scattered X-rays
were detected with a Bicron point-detector after reflection on the
crystal analyzer. The resolution of the diffractometer for a $2
\theta$ -- scan was 0.002 $^o$ for the value of $2 \theta$ and the
FWHM of the direct beam was equal to 0.005 ${^o}$ in $2 \theta$.
For temperature regulation we used a home-made oven and a
computer-driven temperature controller.

Samples were contained in flat glass capillaries of thickness 0.1
mm (Vitro Com Inc., Mountain Lakes, New Jersey, U.S.A.).
Capillaries were filled with the hexagonal phase at room
temperature by suction using a vacuum pump, as described in detail
in reference \cite{imperor}. This procedure allows us to obtain
very well aligned samples, the surfactant cylinders being oriented
along the capillary long axis by the flow.

In order to investigate the structure of the isotropic phase,
X-ray scattering experiments were also performed at the
Laboratoire de Physique des Solides using a rotating anode set-up
(Cu $K_{\alpha}$, $\lambda$ = 0.154 nm). The X-ray beam delivered
by the anode is punctually focused by two perpendicular curved
mirrors coated with a 60 nm nickel layer \cite{imperor}. The
mirrors cut the high energy radiation issued from the anode and a
20 $\mu$m nickel foil filters the $K_{\beta}$ emission line. The
beam size on the sample was $ 0.5  \times 0.5 \, \mbox{mm} ^2$.
The X-ray intensity at the sample level is about $10^7$
\mbox{photons/s mm$^2$}. The scattered X-rays were detected on
imaging plates and the sample -- detection distance was 30 cm.
Exposure times were typically 10 hours in the liquid isotropic
phase.

\section{Results}

\subsection{Diffusion}

\begin{figure}[htbp]
\includegraphics[width=12cm]{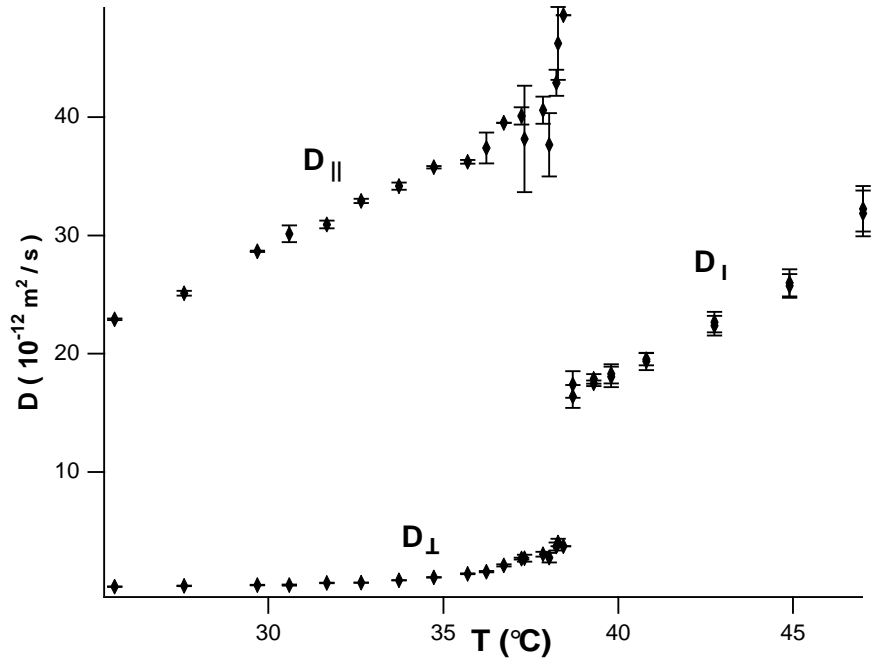}
\caption{\protect \small Diffusion coefficients in the hexagonal
and isotropic phases : $D_{\|}$, $D_{\bot}$ and $D_I$.}
\label{fig:dpoi}
\end{figure}

\begin{figure}[htbp]
\includegraphics[width=12cm]{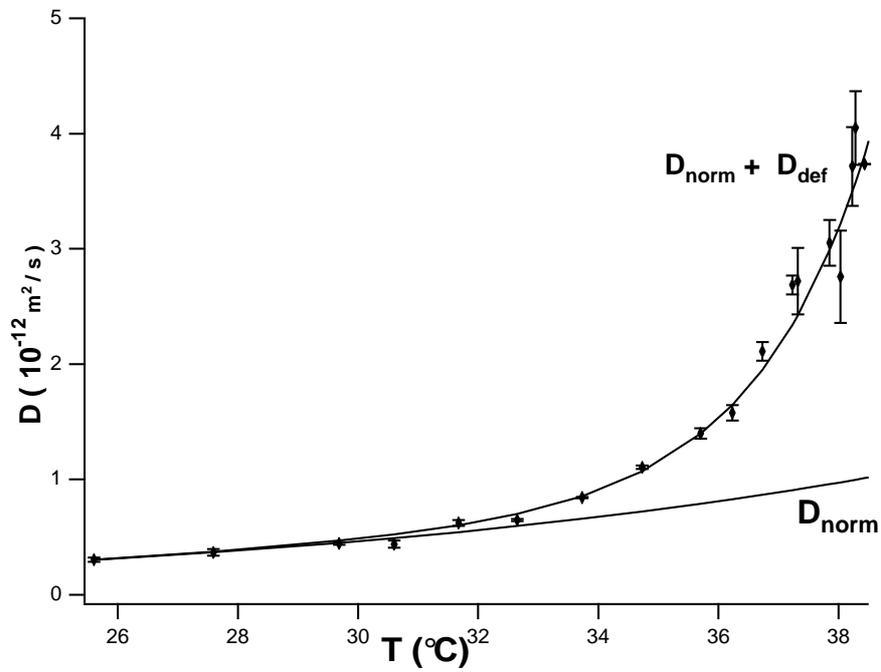}
\caption{\protect \small $D_{\bot}$ in the hexagonal phase. Error
bars are obtained by an average over three different measures. The
straight line represents the extrapolation of low-temperature
behaviour. The solid curve is an exponential fit (see text).}
\label{fig:ddef}
\end{figure}

The results for the hexagonal-isotropic transition are presented
in figures \ref{fig:dpoi} and \ref{fig:ddef}. Far below the
transition, both $D_{\|}$ and $D_{\bot}$ in the hexagonal phase
follow Arrhenius laws, with activation energies of 0.35 eV and
0.75 eV respectively. Starting at $T \sim T_h - 5 \dgr$, the
behaviour of $D_{\bot}$ changes. Its value increases rapidly,
departing from the activation law and reaches $4.1 \, 10^{-12}
\mbox{ m}^2/\mbox{s}$ at the transition temperature $T_h = 38.60
\dgr$. The behaviour of $D_I$ (in the isotropic phase) also fits
with an Arrhenius law, giving an activation energy of 0.65 eV.

Close to the liquid crystal transition temperature we attribute
the increase in $D_{\bot}$ to the proliferation of defects
connecting the cylinders and providing passage for the tracer
molecules. The next section presents a detailed and quantitative
discussion of this phenomenon.

In order to confirm this conclusion we measured the diffusion
coefficient at the cubic-isotropic phase transition. This type of
experiment has already been performed by Monduzzi and coworkers
\cite{monduzzi} for the self-diffusion coefficient of surfactant
in the ionic system CPyCl/NaSal/D$_2$O. Using PFG-NMR, they found
little or no difference between the diffusion coefficients in the
cubic and isotropic phases.

\begin{figure}[htbp]
\includegraphics[width=12cm]{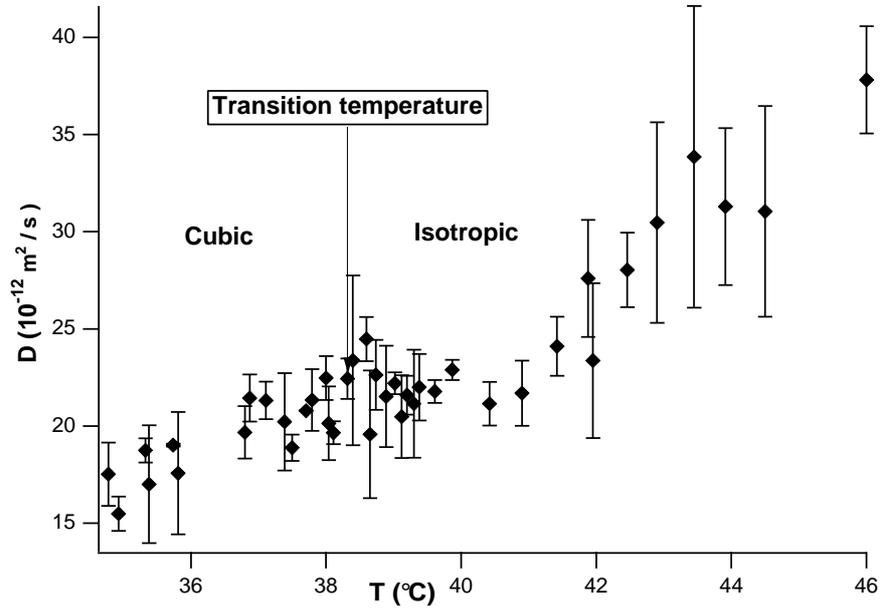}
\caption{\protect \small Diffusion coefficients in the cubic and
isotropic phases, $D_C$ and $D_I$. Error bars are obtained by an
average over three different measures.} \label{fig:dcubic}
\end{figure}

Figure \ref{fig:dcubic} shows our experimental results. The
transition temperature is  $T_c = 38.3 \dgr$. Within experimental
accuracy, no discontinuity was observed at the transition. Since
the diffusion rate is closely related to the local structure of
the phase, we can then conclude that, at least at concentrations
of about 60 \%, the connectivity of the isotropic phase is similar
to that of the bicontinuous cubic one.

\subsection{X-Ray Scattering}

\begin{figure}[htbp]
\includegraphics[width=12cm]{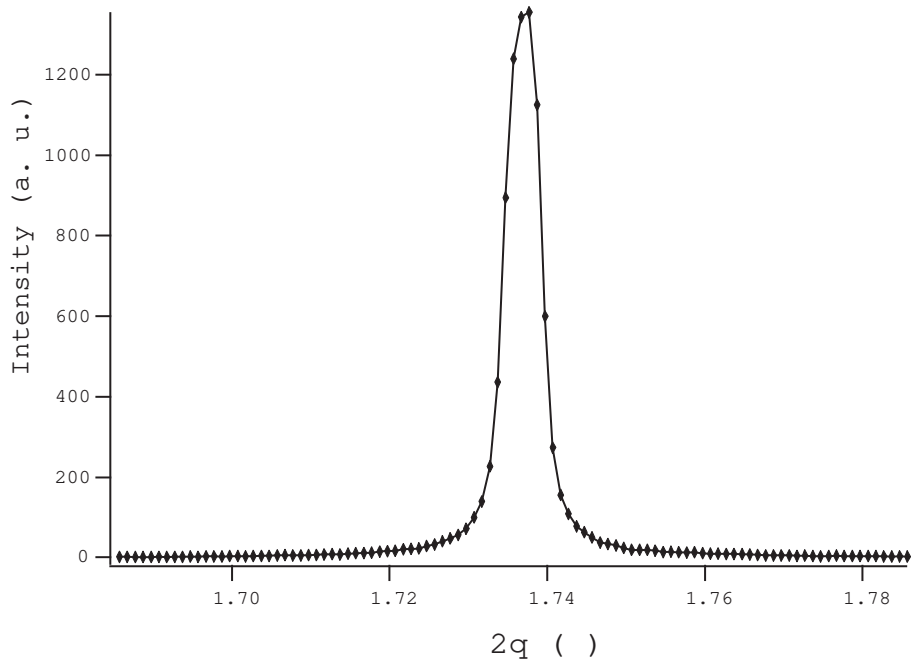}
\caption{\protect \small Profile of the (10) Bragg peak in the
hexagonal phase (limited by the experimental resolution).}
\label{fig:thetascan}
\end{figure}

A well oriented sample of the hexagonal phase (50 w\% of
surfactant) contained in a flat glass capillary of thickness 100
$\mu$m was mounted with a (10) hexagonal Bragg peak in reflection
condition \cite{imperor}. We performed $2 \theta$--scans of this
peak at different temperatures. One such scan is shown in figure
\ref{fig:thetascan}. The value of the hexagonal lattice parameter
$a$ at a given temperature is obtained from the position $2 \theta
_{\rm max}$ of the maximum of the diffraction peak :

\beq
\label{parameter} a=\frac{2}{\sqrt{3}} \, d_{\rm 10} = \frac{2}{\sqrt{3}} \,
\frac{\lambda}{2 \sin (\theta _{\rm max})} =
\frac{\lambda}{\sqrt{3} \sin (\theta _{\rm max})}
\eeq

\begin{figure}[htbp]
\includegraphics[width=12cm]{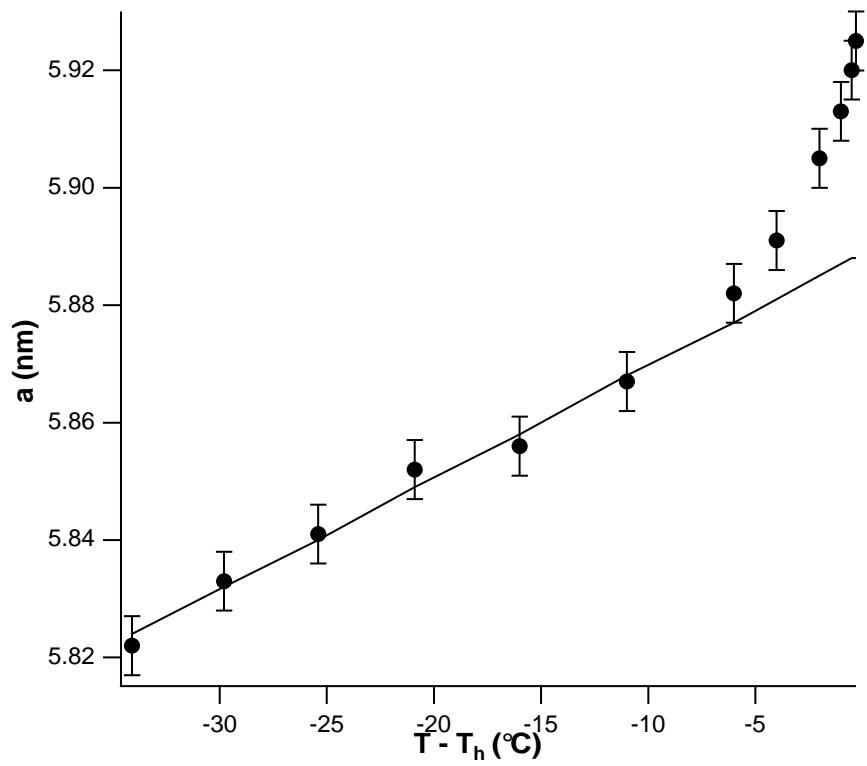}
\caption{\protect \small Lattice parameter $a$ vs. temperature in
the hexagonal phase. The line is the extrapolation of
low-temperature behaviour. } \label{fig:atemp}
\end{figure}

Thanks to the high resolution of the diffractometer, we were able
to detect very small changes (down to 0.002 $^o$) of the value of
$2 \theta _{\rm max}$ versus temperature. The results are plotted
in figure \ref{fig:atemp}. Far below the transition to the
isotropic liquid phase, $a$ increases linearly with temperature :

\beq
\label{alin} a=a_{\rm lin}(T) = (5.9 - 0.0019 (T_h-T)) {\rm nm}
\eeq

\noindent The value of the thermal expansion coefficient is $\ds
\frac{1}{a} \frac{{\rm d}a}{{\rm d}T}=3.2 \, 10^{-4} {\rm
K}^{-1}$. Its positive sign accounts for the observation of the
zig-zag structure on cooling the hexagonal phase \cite{oswald2}.
Near the phase transition, one observes (figure \ref{fig:atemp}) a
deviation from the linear temperature dependence, with an
additional increase in $a$, $\Delta a = a - a_{\rm lin}$. This
effect is observed in the range $(T_h-T) < 5 \dgr$, in good
agreement with birefringence and NMR observations \cite{sallen1}
and the diffusion results.

\begin{figure}[htbp]
\includegraphics[width=12cm]{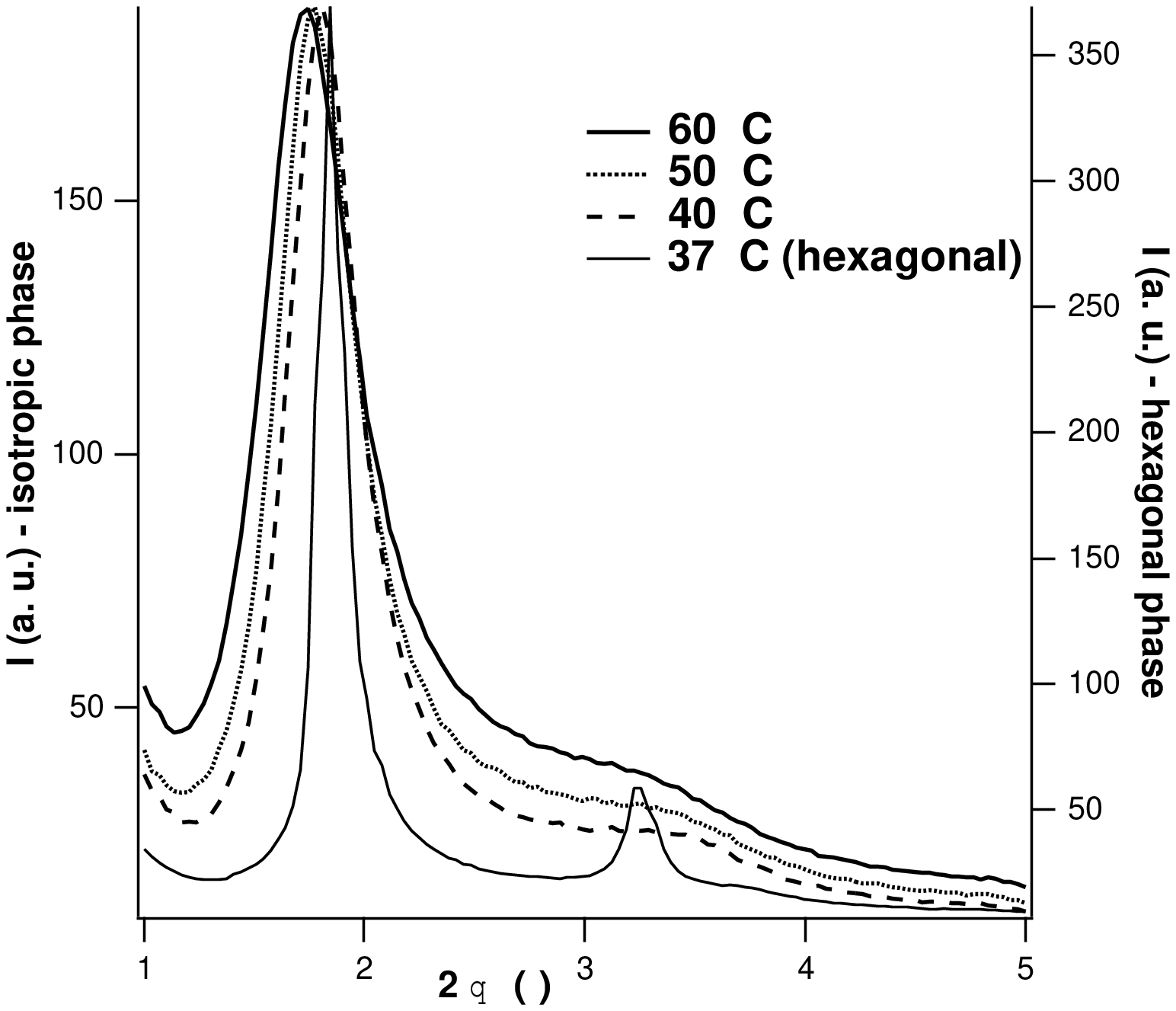}
\caption{\protect \small X-ray spectrum in the isotropic phase. A
scattering profile in the hexagonal phase is shown for comparison
(note the difference in signal intensity).} \label{fig:waxs}
\end{figure}

We performed X-ray scattering measurements on the isotropic phase
at the same surfactant concentration (figure \ref{fig:waxs}). For
comparison, we also plotted the first two Bragg peaks of the
hexagonal phase. It is possible to notice that there is
practically no shift of the maximum scattering vector $q_0$ upon
crossing the transition temperature (compare the curves for 37
$\dgr$ and 40 $\dgr$). This indicates that there is little change
in the local structure of the system. Moreover, the fact that the
peak in the isotropic phase is narrow and that a wide shoulder is
detectable (corresponding to the second Bragg peak, in position
$\sqrt{3}$ as compared to the first), suggests that local order is
preserved, up to a distance $d$ that can be estimated as :

\beq
\label{localorder} d \sim a \frac{q_0}{\Delta q} \sim 4 a
\eeq

\noindent where $\Delta q$ is the width of the peak (in the
isotropic phase) from which we subtract the experimental
resolution, taken as roughly equal to the width of the Bragg peak
in the hexagonal phase \cite{resolution}. The positional order is
maintained up to fourth-neighbours. This should not surprise us,
since we are investigating a very concentrated system.

\section{Discussion}

In this section, we try to relate the pretransitional evolution of
$D_{\bot}$ to the appearance of structural defects. Indeed, NMR
and birefringence data \cite{sallen1} show that, in the same
temperature range ($T_h - 5 \dgr$ up to $T_h$), the hexagonal
order weakens when approaching the transition. This feature can be
explained by considering that an increasing fraction of surfactant
molecules belongs to the defects. These defects may consist either
in a  fragmentation of the columns, which become spherically
capped cylinders, or in bridges connecting neighboring columns (
figure \ref{fig:defects}), as discussed in the Introduction (see
also reference \cite{sallen1}). A possible structure for the
latter type  of defects has been proposed in terms of Karcher
surfaces connecting columns three by three \cite{clerc,karcher}.

\subsection{Diffusion}

\begin{figure}[htbp]
\includegraphics[width=12cm]{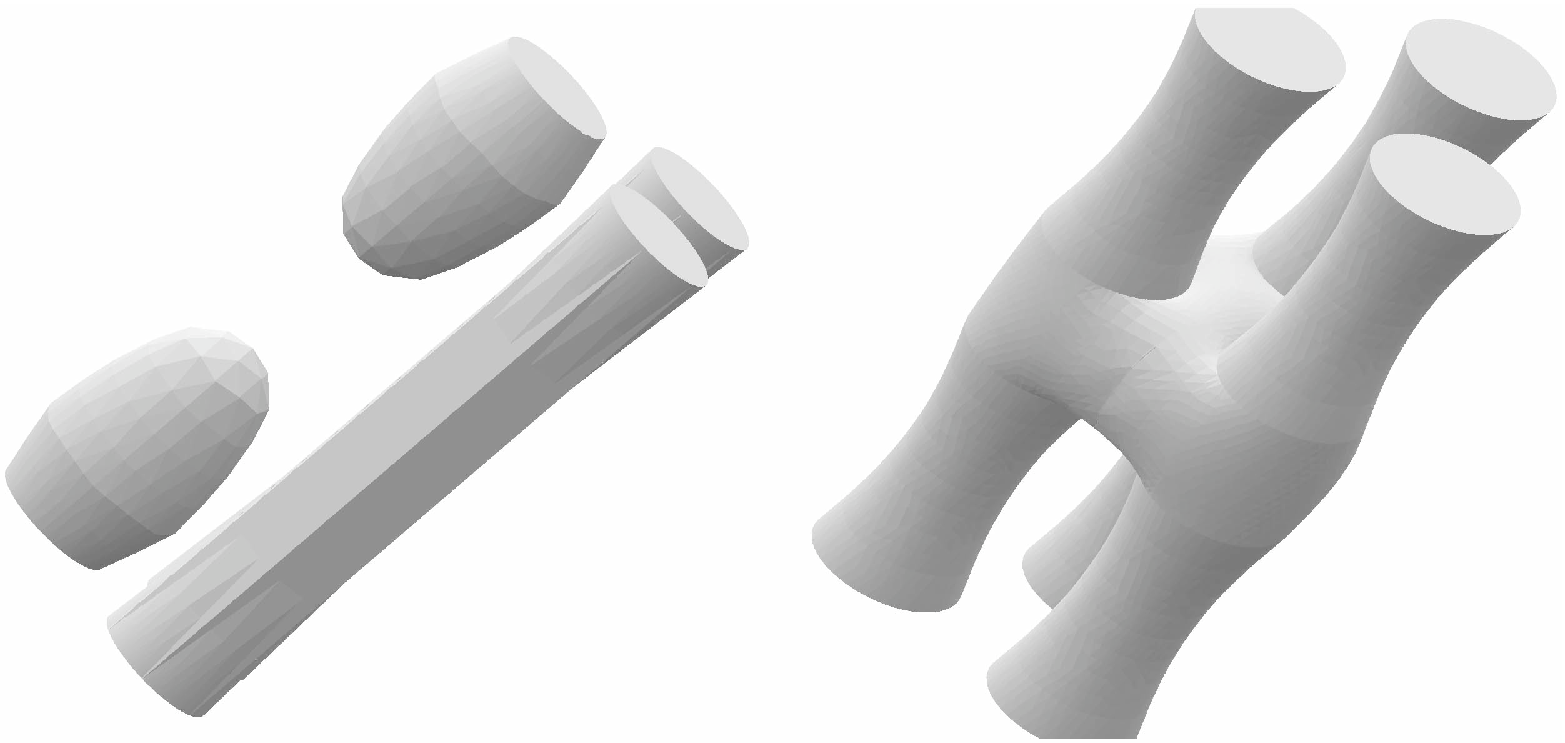}
\caption{\protect \small Possible structures for defects in the
hexagonal phase : a) - capped cylinders; b) - bridges.}
\label{fig:defects}
\end{figure}

It is clear that connections between cylinders can provide a
passage for the tracer molecules, thus leading to the observed
increase in $D_{\bot}$. The presence of defects of the first type
(capped cylinders) can be ruled out, since it would bring about a
decrease in $D_{\|}$, which does not appear in our measurements.
We will assume in the following that the only defects present are
of the `bridge' type (figure \ref{fig:defects} -- b). This
conclusion is coherent with the fact that the curvature of the
aggregates diminishes with increasing temperature (due to the
decreasing hydration of the nonionic polar groups
\cite{israelachvili,puvvada}) which clearly favours the merging of
the cylinders over their breaking up.

Let us now estimate the density of defects $n_{\rm def}$ (number
of defects per unit volume). In order to do so, we must quantify
their role in transverse diffusion by relating $n_{\rm def}$ to
$D_{\rm def}=D_{\bot}-D_{\rm norm}$ (where $D_{\rm norm}$ is the
``normal'' behaviour of $D_{\bot}$, extrapolated from low
temperature). We shall denote by $a \simeq 6 \mbox{ nm}$ (see
figure \ref{fig:atemp}) the parameter of the hexagonal lattice and
let $L$ be the average distance between connections along a
cylinder. The adimensional parameter $x = a / L$ provides a
quantitative measure of the defect density, since $n_{\rm
def}=\frac{2}{3\sqrt{3}a^3} x$ \cite{density}. We shall see in the
following that the diffusion contribution of the defects also
depends on their lifetime, $\tau$. Throughout the discussion, we
will consider only the type of defects depicted in figure
\ref{fig:defects} -- b, connecting three neighbouring cylinders.

In the case where the lifetime of the defects is very short
compared to the characteristic diffusion time of the molecules,
$D_{\rm def}$ can be evaluated analytically. The molecule has a
constant probability $x$ to encounter a defect. Once a defect is
reached, the molecule can either cross it, or continue to move
along the cylinder, with equal probability, so that it will spend
(on average) $x/2$ of its time crossing defects and $1-x/2$
travelling along the columns. Denoting by $z$ the axis of the
cylinders and by {\boldmath{$\rho $}} the position vector in the
plane orthogonal to $z$, the presence probability for a particle
starting a random walk at the origin is (time is given in units of
$t_0$, the elementary step) :

\beq
\label{proba} P({\mbox{\boldmath{$\rho $}}},z,t) = C(t) \exp \left ( \frac{-z^2}{2 a^2 (1-x/2)t} \right ) \exp \left ( \frac{-{\mbox{\boldmath{$\rho $}}}^2}{(4/3) a^2 (x/2) t} \right )
\eeq

\noindent where $C(t)$ is a normalisation factor depending only on
the time; the numerical factors 2 and 4/3 are the ones for random
walks respectively in 1D and in 2D on a hexagonal network. Keeping
in mind that we are interested not in the density along
{\boldmath{$\rho $}}, but in its projection on the fixed plane of
the sample (which gives an additional factor $1/ \sqrt{2}$ ) we
can estimate the ratio $\ds r = D_{\rm def} / {D_{\|}}$ as being :

\beq
\label{ddef} r = \frac{2}{3 \sqrt{2}} \frac{x/2}{1-x/2}
\eeq

In order to check the validity of formula (\ref{ddef}), and to
investigate the influence of the defect lifetime, we have
performed Monte Carlo numerical simulations. We generate random
walks on a 3D lattice which reproduces the hexagonal structure.
The elementary RW jump is equal to the lattice parameter. In the
absence of defects, the particle only diffuses along axis $z$.
Defects connecting three nearest-neighbour cylinders are
introduced in order to allow transverse diffusion. When a defect
is encountered, the particle has a probability of 1/2 to remain on
the same cylinder and equal probabilities 1/4 to jump on one or
the other of the two connected neighbours.

One elementary RW step defines the unit time in the simulation.
Defects have a lifetime $\tau$ which means that, every $\tau$
elementary steps, all the defects are erased and replaced again at
random.

Each RW has $10^4$ steps and the probability distribution is averaged over $10^4$ RWs.

\begin{figure}[htbp]
\includegraphics[width=12cm]{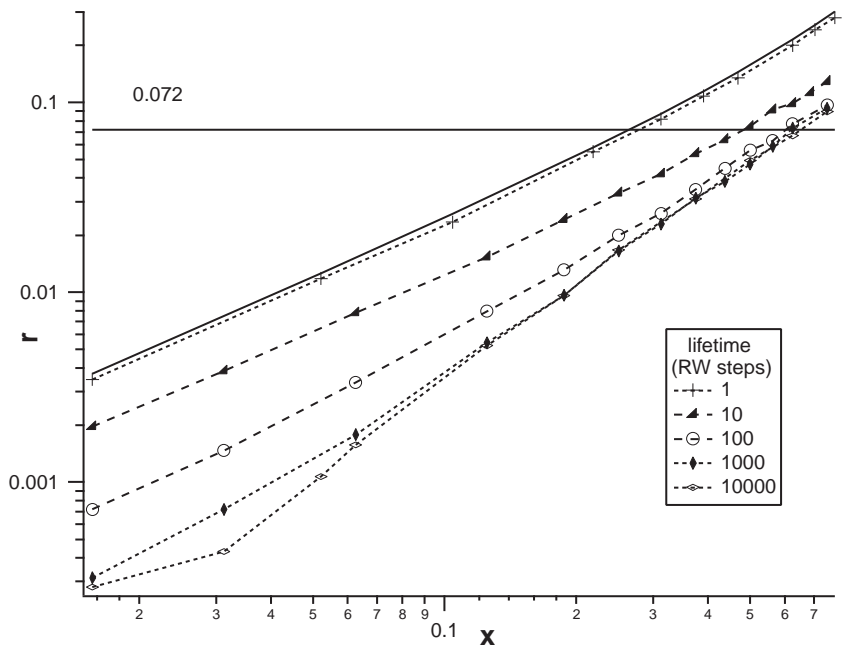}
\caption{\protect \small $\ds r = D_{\rm def} / {D_{\|}}$ plotted
against $x = a/L$ for different defect lifetimes $\tau$. }
\label{fig:simul}
\end{figure}

The statistics of particle positions gives us the diffusion
coefficients both the parallel (along the $z$ axis) and
transversal diffusion coefficients. Their ratio $\ds r = D_{\rm
def} / {D_{\|}}$ is plotted against $x = a/L$ (figure
\ref{fig:simul}) for different lifetimes $\tau$.

For $\tau = 1$, the defect configuration should be changed at
every RW step and the simulation would require too much time.
However, since the defect positions are completely decorrelated
from one step to the other, the solution adopted was to give each
particle a constant probability $x$ of encountering a defect. We
see that the dependence $r(x)$ is very well described by equation
(\ref{ddef}) (solid line in figure \ref{fig:simul}).

It is clear from figure \ref{fig:simul} that increasing the
stability of the defects lessens their efficiency for tracer
diffusion or, in other words, more defects are needed to obtain a
given value of $r$ when $\tau$ becomes larger : if for $\tau = 1$
a density $x \simeq 0.25$ is needed in order to reach the value
measured just before the transition to the isotropic phase $r =
D_{\rm def} / D_{\|} = 0.072$, for $\tau = 10$ one already has $x
\simeq 0.5$.

The dependence of $r(x)$ also changes its character with
increasing $\tau$ : for $\tau = 1$, $r \sim x$ (to leading order),
while for frozen-in defects ($\tau = 10^4$), $r \sim x^2$. This is
because, in order to jump from a cylinder to the next one at a
distance $a$, the molecule has to diffuse along $z$ for a distance
$L$ :

\beq
t_{\rm jump} \sim \frac{a^2}{D_{\rm def}} \sim \frac{L^2}{{D_{\|}}} \Longrightarrow r = \frac{D_{\rm def}}{{D_{\|}}} \sim \frac{a^2}{L^2} = x^2
\eeq

Since the defects are at equilibrium, their density $n_{\rm def}$ is given by a Boltzmann factor

\beq
\label{boltzmann} n_{\rm def}=n_0 \exp \left ( \frac{-E_{\rm def}}{k_B T} \right )
\eeq

\noindent where $E_{\rm def}$ is the energy of the defect as
compared to the perfectly ordered structure. The fact that the
number of defects increases abruptly close to the transition
temperature means that their energy decreases (we will neglect the
increase in thermal energy : $k_B T \simeq k_B T_h$ in the
vicinity of the transition). The simplest assumption is that of a
linear behaviour :

\beq
\label{edef} E_{\rm def} = \alpha (T_h -T) + E_0
\eeq

The birefringence and NMR measurements \cite{sallen1} give (by
fitting $n_{\rm def}(T)$) a value $\alpha = 225 \, k_B$; our fit
for $D_{\rm def} (T)$ (see figure \ref{fig:ddef}) gives $\alpha =
180 \, k_B $. We can thus consider that $D_{\rm def} \sim n_{\rm
def}$, which supports our hypothesis that the defects have very
short lifetimes. It is then justified to use (\ref{ddef}) for
evaluating the mean distance between connections at the transition
temperature $L_0$ : $L_0 \simeq 4 a \simeq 25 \, \mbox{nm}$, in
fairly good agreement with results of reference \cite{sallen1},
which gives $L_0 \simeq \mbox{17 nm}$. An  important consequence
is that the density of defects in the hexagonal phase is large at
the transition; the picture of the hexagonal phase as being formed
of infinitely long parallel columns is no longer accurate in these
conditions.

We can thus infer that the isotropic phase above the hexagonal mesophase has a highly connected structure.
\subsection{X-ray scattering}

The additional increase in lattice parameter $\Delta a$ (figure
\ref{fig:atemp}) can also be explained by the appearance of
connections close to the transition. In this paragraph, we derive
a very simple relation between $\Delta a$ and the fraction $f$ of
surfactant molecules involved in the connections.

\begin{figure}[htbp]
\includegraphics[width=12cm]{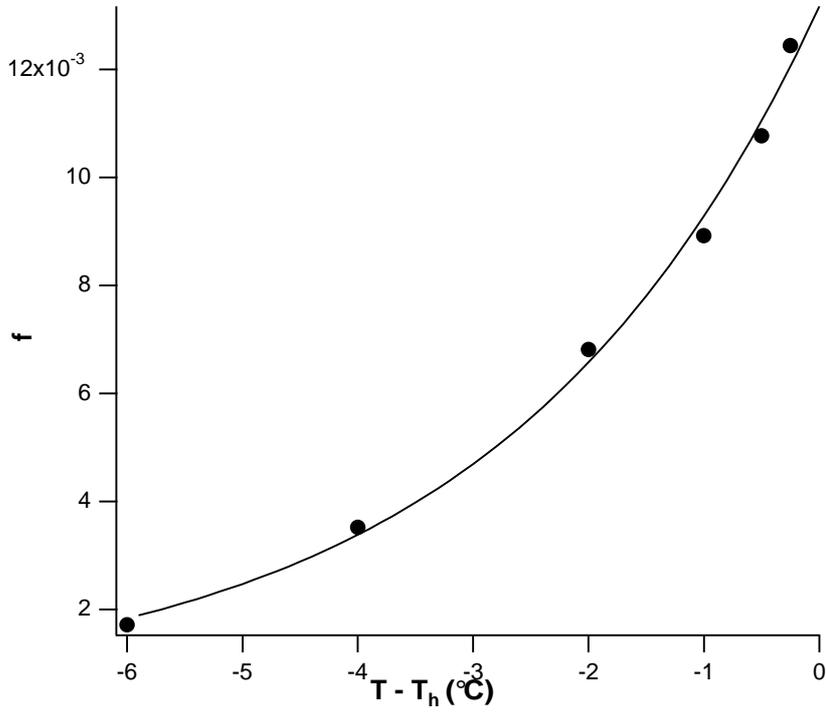}
\caption{\protect \small  The fraction $f$ of molecules present in
the defects as a function of temperature. Line is an exponential
fit (see text).} \label{fig:ftemp}
\end{figure}

 The X-ray peak provides a measure for the parameter of the hexagonal lattice formed by the surfactant
cylinders. If a fraction $f$ of molecules is involved in
transversal junctions between cylinders, a fraction $(1-f)$ of
molecules are still inside the cylinders. Mass conservation then
demands that in a given volume, the total number of cylinders is
divided by $1/(1-f)$ and the lattice parameter is multiplied by a
factor $\ds \frac{1}{\sqrt{1-f}}$ (because the cylinders form a 2D
hexagonal array). The value of $f$ as a function of temperature is
given by the relation :

\beq
\label{afuncf} a(T)=\frac{a_{\rm lin} (T)}{\sqrt{1-f}} \Longrightarrow f = 1 -  \left [ \frac{a_{\rm lin} (T)}{a(T)} \right ] ^2
\eeq

\noindent where the value $a_{\rm lin}(T)$ is extrapolated from
the low-temperature behaviour in the region near the transition
(solid line in \ref{fig:atemp}). From this relation we obtain
$f(T)$, which fits well with an exponential law, as plotted in
figure \ref{fig:ftemp}.

The fact that we find an exponential law for $f(T)$ for $0 < T_h -
T < 5 \dgr$ is in agreement with the tracer diffusion results and
the NMR and birefringence experiments \cite{sallen1}. However, we
obtain $f(T_h) = 1.3 \%$, much less than the value of $8 \%$
estimated in reference \cite{sallen1} (and a value of the
parameter $\alpha \simeq 100 k_B T$, about half of that given by
the birefringence experiment). This is not very surprising, since
it is reasonable to assume that the connections induce an elastic
deformation of the hexagonal lattice, which will also affect the
value of the average lattice parameter $a$ as measured by X-ray
scattering. If such a contribution tends to reduce locally the
value of $a$ (the cylinders getting closer together), then the
simple relation (\ref{afuncf}) underestimates the value of $f$.

\section{Conclusion}

The investigation of the \formeau system by means of diffusion
coefficients measurements and X-ray scattering allowed us to
obtain a clearer image of its isotropic phase for high surfactant
concentration (50--60 \%). The data was mainly obtained by
studying the pretransitional effects that appear in the hexagonal
and bicontinuous cubic mesophases close to the transition towards
the isotropic phase.

The increase of the diffusion coefficient across the cylinders
($D_{\bot}$) in the hexagonal phase for  a hydrophobic fluorescent
dye proves that very mobile defects, consisting in connections
between the cylinders, appear close to the transition.

This information is corroborated by the anomalous increase of the
lattice parameter $a$ in the same temperature range. The X-ray
scattering of the isotropic phase shows a well-defined and fairly
narrow peak corresponding to the first Bragg peak in hexagonal
phase.

No detectable jump in diffusion coefficient occurred at
thetransition between the cubic and isotropic phases, showing that
the surfactant aggregates in the two phases are very similar.

We can therefore conclude that, in the concentration range that we
investigated, the isotropic phase of the \formeau system is
probably composed of very long surfactant cylinders locally
preserving the hexagonal order (even though long-range order is
lost), forming a highly connected and rapidly fluctuating
structure.

{\bf Acknowledgements.} M. Gailhanou (LURE, beamline H10) is
warmly thanked for taking part in the synchrotron X-ray scattering
experiments. We acknowledge fruitful discussions with Robert
Ho{\l}yst. D. C. gratefully acknowledges financial support from
the research group `Liquid crystals in confined geometries' during
his stay in Orsay.

\newpage
\begin{center}
{\large\bf FIGURES}
\end{center}

\begin{thebibliography}{99}
\bibitem{resolution} In the hexagonal phase, the Bragg peak should be a Dirac $\delta$ function (infinite lattice). Its width is only given by the experimental resolution. We must therefore subtract this width in order to assess the order in the isotropic phase. The remaining $\Delta q$ gives then a direct measure of local order in the system : if $d$ is the distance over which hexagonal order is preserved, $\Delta q \sim d^{-1}$, as $q_0 \sim a^{-1}$; hence, the relation \ref{localorder}.
\end{thebibliography}
\end{document}